\documentclass[letterpaper, 10 pt, conference]{ieeeconf}

\IEEEoverridecommandlockouts                              

\usepackage{amsthm}
\usepackage{cite}
\usepackage{amsfonts}
\usepackage{algorithmic}
\usepackage{graphicx}
\usepackage{textcomp}
\usepackage{xcolor}
\usepackage{amsmath}
\usepackage{amssymb}
\usepackage{adjustbox}

\usepackage{subcaption}

\theoremstyle{definition}
\newtheorem{definition}{Definition}

\usepackage{makecell}

\title{\Large \bf
Improving the Region of Attraction of a Multi-rotor UAV by Estimating Unknown Disturbances}

\author{Sachithra Atapattu$^{1}$, Oscar De Silva$^{1}$, Thumeera R Wanasinghe$^{1}$, George K I Mann$^{1}$,  and Raymond G Gosine$^{1}$
\thanks{*This research is funded by Natural Sciences and Engineering
Research Council of Canada (NSERC) Discovery grants and Memorial University of Newfoundland.}
\thanks{$^{1}$Sachithra Atapattu, Oscar De Silva, George K I Mann, Thumeera R Wanasinghe, Raymond G Gosine are with Faculty of  Engineering, 
        Memorial University of Newfoundland, St. John's, NL A1C5S7 Canada
        {\tt\small shatapattu@mun.ca}, {\tt\small oscar.desilva@mun.ca}, {\tt\small gmann@mun.ca}, {\tt\small thumeerawa@mun.ca}, {\tt\small rgosine@mun.ca}}%
}

\begin{document}

\maketitle
\thispagestyle{empty}
\pagestyle{empty}

\begin{abstract}
This study presents a machine learning-aided approach to accurately estimate the region of attraction (ROA) of a multi-rotor unmanned aerial vehicle (UAV) controlled using a linear quadratic regulator (LQR) controller. Conventional ROA estimation approaches rely on a nominal dynamic model for ROA calculation, leading to inaccurate estimation due to unknown dynamics and disturbances associated with the physical system. To address this issue, our study utilizes a neural network to predict these unknown disturbances of a planar quadrotor. The nominal model integrated with the learned disturbances is then employed to calculate the ROA of the planer quadrotor using a graphical technique. The estimated ROA is then compared with the ROA calculated using Lyapunov analysis and the graphical approach without incorporating the learned disturbances. The results illustrated that the proposed method provides a more accurate estimation of the ROA, while the conventional Lyapunov-based estimation tends to be more conservative.

\end{abstract}

\section{INTRODUCTION}

Multi-rotor unmanned aerial vehicles (UAVs) are inherently unstable. Due to that, their operational safety depends heavily on the effectiveness of the control system \cite{DNN_ICRA_cite}. The safety of a multi-rotor UAV can typically be maintained by operating close to their hovering points, but this can limit their agility and, consequently, the efficiency of operations. Moreover, multi-rotor UAVs face significant challenges when operating under external disturbances like strong winds, which are frequent in outdoor settings.

Control strategies for multi-rotor UAVs vary widely, ranging from traditional feedback mechanisms like Proportional-Integral-Derivative (PID) controllers to more sophisticated, machine learning-based systems such as neural controllers \cite{neural_controller_cite}. Although PID controllers have simplicity and ease of implementation, they often struggle with nonlinear systems or those with complex dynamics. On the other hand, neural network-based controllers excel in managing non-linearities and complex decision-making but typically require specialized hardware and involve substantial computational demands, which can extend processing times. In contrast, the Linear Quadratic Regulator (LQR) controller stands out as a well-established option for various dynamic systems, demonstrating competence in handling nonlinear and complex dynamics.  Additionally, the computational complexity of the LQR controller is noticeably lower than neural controllers. 
Recent advancements in controller performance include the integration of machine learning techniques such as neural networks, Gaussian process regression, and parametric regression methods \cite{DNN_ICRA_cite, sachithra_cite, learn_fast_cite, safe_learning_ROA_cite} to learn unknown dynamics of multi-rotor UAV systems. These approaches have been shown to enhance the robustness of controllers.

To ensure the safe operation of multi-rotor UAVs under a broader spectrum of conditions, such as strong winds and model uncertainties caused by payloads, it is crucial to assess the controllers' capability to manage deviations from stability points. The region of attraction (ROA) for multi-rotor UAVs defines the complete set of initial states—including positions, velocities, and orientations—from which the multi-rotor UAV's control system can reliably return it to a desired equilibrium state or trajectory over time \cite{convex_cite}. This concept of ROA is vital for gauging the multi-rotor UAV's resilience in stabilizing itself from varied start conditions or deviations from desired trajectory states. A well-defined ROA that covers the whole set of operating conditions allows multi-rotor UAVs to safely perform their tasks when contending with unknown dynamics and disturbances. The stability within the space defined by the ROA is guaranteed; hence, safe trajectories can be planned in proximity to people or valuable assets, allowing the multi-rotor UAV to recover and proceed with its mission even under unforeseen conditions.

Despite LQR-based control strategies have been used for UAV navigation \cite{sachithra_cite} and their competitive advantages over simple PID controllers or neural controllers, there is a notable gap in the literature regarding the integration of machine learning techniques to estimate unknown dynamics for improving the ROA in conjunction with LQR controllers. Our study addresses this gap by adopting a control strategy that effectively manages nonlinear dynamics through a learned dynamic model. By maintaining constant controller gains, we utilize the refined dynamic model to calculate thrust inputs. This method is poised to significantly expand the ROA, thereby increasing the system's adaptability without altering the control strategy. This ensures safety while adapting to dynamic disturbances.

The contributions of this work can be highlighted as follows.
\begin{itemize}
    \item Improving the dynamic model of a multi-rotor UAV using machine learning techniques.
    \item Evaluation of improvement of ROA by incorporating learned dynamics.
    \item Comparison of ROA estimations from Lyapunov function and graphical method.
\end{itemize}




\section{BACKGROUND}

Recent research has turned to novel methods for improving ROA estimation, leveraging analytical techniques and experimental insights. In this section, findings from several key studies are integrated to present a comprehensive review of the context of the improvement of ROA and machine learning techniques. 

Tobenkin et al. \cite{inv_funnels_SOS_cite}, a leading work in the field of ROA, presents numerical methods for computing regions of finite-time invariance (funnels) around solutions of polynomial differential equations. It introduces two main approaches: one uses exact trajectory data to generate time-varying polynomial Lyapunov functions, while the other uses a time-sampling technique to approximate these functions more rapidly. These funnels are verified using Sum-of-Squares (SOS) programming, contributing significantly to the control system's design by ensuring that initial conditions remain within certain bounds, leading to a desired endpoint. The ROA is improved through the formulation and verification of “funnels,” which are regions in the state-space where all trajectories converge to a goal region at a specified time. The methods ensure that even with approximate trajectories, the invariance conditions are maintained. By leveraging time-varying Lyapunov functions and SOS programming, their study can efficiently compute inner approximations of the ROA. The time-sampled approach, in particular, is noted for being significantly faster while achieving nearly identical results to the exact method.

The work presented in \cite{traj_measurements_ROA_cite} offered a technique reliant on trajectory measurements to estimate the ROA, bypassing the need for explicit knowledge of the system's nonlinear vector field. This approach is based on developing a learned Lyapunov function from trajectory data using least absolute deviations and SOS polynomial optimization. The learned Lyapunov functions are then used to determine whether new initial conditions fall within the ROA. The methodology has been validated numerically, showing over 95\% prediction accuracy on test data. Their highly accurate, data-driven approach is suitable for systems with highly complex and hard-to-determine dynamics. However, the use of SOS polynomials and semi-definite programming can be computationally intensive, especially as the state dimension and the degree of the polynomial increase. This might limit real-time applications on UAVs with constrained computational resources. In addition, this method might yield conservative estimates of the ROA, particularly if a conservative approach is used to define the level sets of the Lyapunov function.

Authors in \cite{lyap_NN_cite} introduce a novel neural network-based Lyapunov function to certify safety and improve the ROA estimation in nonlinear dynamical systems. This method enhances safe learning capabilities by adapting the shape of the ROA based on real-time system inputs and outputs, overcoming limitations of traditional methods that rely on fixed or predefined ROA shapes. The adaptive feature allows for more effective safe exploration in uncertain environments, crucial for applications like autonomous robotics. However, despite these advancements, the approach's reliance on accurate data sampling and the complexity of neural network training might limit its applicability in highly dynamic or unpredictable settings, where data may not sufficiently represent all system states.

\cite{piecewise_lyap_cite} presents a novel method for estimating the ROA for switched nonlinear systems through the use of piecewise polynomial Lyapunov functions. This approach aims to overcome the limitations of using a common Lyapunov function, which can be conservative when systems do not share a single Lyapunov function throughout the trajectory. The technique involves finding separate Lyapunov functions for each subsystem and ensuring their continuity during transitions. This enables a less conservative estimation of the ROA compared to common Lyapunov function methods. The proposed method utilizes SOS programming to verify that a given piecewise Lyapunov function's sublevel set is within the actual ROA. This employs a combination of Square Matrix Representation (SMR) and SOS programming to find the optimal piecewise Lyapunov function. 

The work presented in \cite{preching_glider_cite} leverages LQR and Rapidly-exploring Random Trees (RRT) to generate a controller with verified performance across a range of initial conditions, applying time-varying Lyapunov functions to ensure safety and stability. It systematically expands the ROA for UAV perching maneuvers. By mapping out a robust network of feasible trajectories and designing localized LQR controllers at each node, the method enhances the stability and reachability from various initial conditions, significantly widening the ROA. However, this paper mainly focuses on dynamic post-stall maneuvers with a fixed-wing UAV, rather than improving the ROA.

Together, these studies emphasize strategies to enhance the ROA for nonlinear dynamic systems. Traditional methods based on Lyapunov analysis tend to yield conservative estimates of the ROA, suggesting that refining Lyapunov analysis could lead to improvements. 

Diverging from these traditional approaches, Berkenkamp et al. \cite{safe_learning_ROA_cite} proposed a method for estimating the ROA for dynamic systems by incorporating machine learning techniques, particularly focusing on ensuring safety during the learning process. Unlike traditional approaches that rely on known system models, this method leverages Gaussian Processes (GPs) to learn from real system data while staying within the true ROA, thereby avoiding safety-critical failures. The process involves using prior knowledge of model uncertainties to guide the learning, ensuring that the learned ROA is accurate and reliable. This approach is particularly useful when exact system dynamics are unknown or partially known, and there is a need to safely explore the system's behavior.

Incorporating machine learning techniques offers opportunities for developing controllers that can adjust to dynamic disturbances, while ensuring safety. To the best of our knowledge, there is a noticeable research gap in estimating unknown dynamics to enhance the ROA. This work addresses this gap by improving system dynamics to enhance the ROA.

\section{MATHEMATICAL FORMULATION}

\subsection{Planar Quadrotor Dynamics}

\begin{figure}[!b]
  \centering
  \includegraphics[height=1.4in]{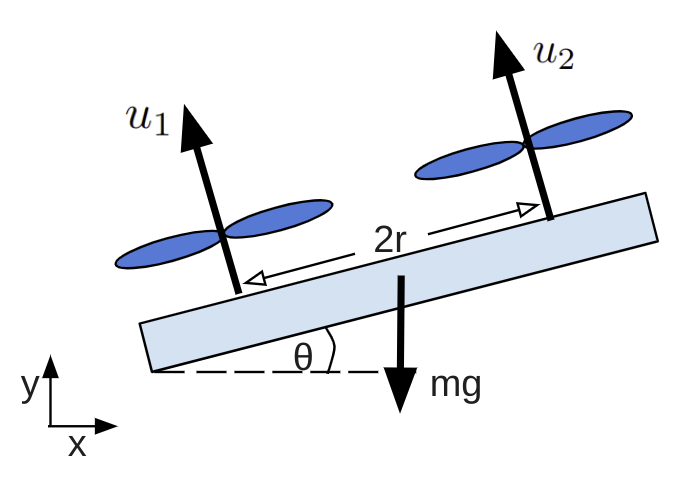}\\
  \caption{Planar quadrotor system}
\label{planar_quad}
\end{figure}

The planar quadrotor control system in Fig. \ref{planar_quad} is defined using its states and control inputs as follows.

\begin{equation}
    \mathbf{x} = 
    \begin{bmatrix}
        x, v_x, y, v_y, \theta, \omega
    \end{bmatrix}^T
\label{states}
\end{equation} and

\begin{equation}
    \mathbf{u} = 
    \begin{bmatrix}
        u_1, u_2
    \end{bmatrix}^T,
\label{inputs}
\end{equation} where $x$ and $y$ are the position coordinates in the $x-y$ plane, $v_x$ and $v_y$ are the velocities along the $x$ and $y$ directions, respectively, $\theta$ is the angle of rotation relative to the $x$-axis, and $\omega$ is the angular velocity.

The governing dynamic equations are shown in \eqref{dynamics}. In these equations, $u_1$ and $u_2$ are the thrusts produced by each propeller, while $g$ represents the gravitational force acting downward along the negative-$y$ axis. The system’s physical parameters—mass ($m$), rotor distance ($r$), and inertia ($I$)—are extracted from \cite{planar_param_cite}, and are set to $m=0.486\,\mathrm{kg}$, $r=0.25\,\mathrm{m}$, and $I=0.00383\,\mathrm{kgm^2}$.

Additionally, as given in \eqref{known_unknown}, a disturbance model, $\pi(\mathbf{x,u})$, is integrated into the system dynamics, $f(\mathbf{x,u})$, to account for external influences and perturbations that affect the quadrotor’s behavior.

\begin{equation}
    \dot{\mathbf{x}} = 
    \begin{bmatrix}
        v_x \\
        \frac{-(u_1+u_2)}{m}\sin(\theta) \\
        v_y \\
        \frac{(u_1+u_2)}{m}\cos(\theta)-g \\
        \omega \\
        \frac{r(u_1-u_2)}{I}
    \end{bmatrix}
\label{dynamics}
\end{equation}

\begin{equation}
  \mathbf{\dot{x}} = 
  \underbrace{f(\mathbf{x,u})}_\text{Dynamic model} 
   + 
  \underbrace{\pi(\mathbf{x,u})}_\text{Disturbance model} 
\label{known_unknown}
\end{equation}

The disturbance model outlined in \eqref{disturbance} has three components, disturbance due to change in mass, disturbance due to rotor aerodynamic drag, and disturbance due to fuselage aerodynamic drag  \cite{hakim_cite}. These disturbances were incorporated into the accelerations, $\dot{v}_x$ and $\dot{v}_y$, where the controller inputs impact the system within the LQR framework.

\begin{equation}
  \pi(\mathbf{x,u}) = -\frac{\mathbf{u}\times\Delta m}{m \times m_T} - \frac{\mathbf{D_r} {}^b\mathbf{v}}{m_T} - \frac{D_f \mathrm{sign}({}^b\mathbf{v}) {}^b\mathbf{v}^2}{m_T},
\label{disturbance}
\end{equation} 

In this disturbance model, ${}^b\mathbf{v}$ denotes the velocity of the quadcopter relative to its body frame, and $\mathbf{u}=[u_1, u_2]^T$ represents the total thrust generated by the propellers. $\Delta m$ indicates the change in mass, resulting in a total mass of $m_T = m + \Delta m$. The matrix $\mathbf{D_r}$ is the rotor drag coefficient matrix and the constant $D_f$ is the fuselage drag coefficient. Specifically, $\mathbf{D_r}$ is defined as a diagonal matrix with entries $[0.2, 0.4]$, reflecting the rotor drag coefficients, and $D_f = 0.3$ represents the fuselage drag coefficient \cite{hakim_cite}.

These components are essential for accurately modeling the quadrotor's response to external forces and internal changes, such as variations in mass or aerodynamic properties. By incorporating these disturbances into the quadrotor’s acceleration calculations, the LQR controller can more effectively compensate for them, thereby enhancing the stability and handling of the quadrotor under various flight conditions.

\subsection{Controller Model}

A linearized state-space model of the planar quadrotor is employed within the continuous domain to implement the control strategy. This model simplifies the dynamics of the quadrotor into a form that is suitable for LQR control design. The linearization process involves approximating the nonlinear dynamics of the quadrotor around a nominal operating point, in this case, hovering at an equilibrium point.

This linearized state-space model is given in \eqref{LQR_state_space}, and was derived by linearizing its dynamics in \eqref{dynamics} around the equilibrium point, $(0,0)$. The thrust generated by the propellers at this equilibrium was set at $u = [\frac{mg}{2}, \frac{mg}{2}]$. 

\begin{equation}
  \dot{\mathbf{x}} = \mathbf{Ax} + \mathbf{Bu}
\label{LQR_state_space}
\end{equation} where,

\begin{equation*}
  \mathbf{A} = 
  \begin{bmatrix}
    0 & 1 & 0 & 0 &  0 & 0\\
    0 & 0 & 0 & 0 & -g & 0\\
    0 & 0 & 0 & 1 &  0 & 0\\
    0 & 0 & 0 & 0 &  0 & 0\\
    0 & 0 & 0 & 0 &  0 & 1\\
    0 & 0 & 0 & 0 &  0 & 0 
  \end{bmatrix}
\label{LQR_A} ~~~ \mathrm{and} ~~~   \mathbf{B} = 
  \begin{bmatrix}
    0   &   0\\
    0   &   0\\
    0   &   0\\
    1/m &  1/m\\
    0   &   0\\
    r/I & -r/I
  \end{bmatrix}.
\end{equation*}

The LQR controller operates by calculating an optimal control gain, $\mathbf{K}$, with the goal of minimizing the infinite-horizon cost function specified in \eqref{LQR_J}, where $\mathbf{Q}$ and $\mathbf{R}$ denote the state error weight matrix and the control expenditure. This cost function essentially quantifies the performance of the controller by integrating the squared magnitude of state deviations and control inputs over time, weighting them according to predefined criteria. Minimizing this cost function aids in achieving control that balances state stability with control energy expenditure.

\begin{equation}
    \mathbf{J} = \int_{0}^{\infty} [(\mathbf{x})^T\mathbf{Q}(\mathbf{x}) + \mathbf{u}^T\mathbf{Ru}] dt
\label{LQR_J}
\end{equation}

The output of the LQR controller, in this case, the rotor thrust, $\mathbf{u}$, is calculated using the optimized $\mathbf{K}$, and the state error, $\bar{\mathbf{x}}$ as follows.

\begin{equation}
    \mathbf{u} = -\mathbf{K}\bar{\mathbf{x}}; ~~~ \bar{\mathbf{x}} = \mathbf{x} - \mathbf{x}_{des}
\label{LQR_u}
\end{equation} where, $\mathbf{x}_{des}$ is the desired equilibrium state, and $\mathbf{x}$ is the current state.

\subsection{Sensitivity of NN}

The estimated disturbance, $\pi_1(\mathbf{x,u})$ and $\pi_2(\mathbf{x,u})$ is added to $\dot{v}_x$ and $\dot{v}_y$ in the dynamic model defined in \eqref{dynamics}. The LQR state-space matrices, $A$ and $B$ were updated accordingly. In order to calculate the partial derivatives of the neural network's outputs in terms of its inputs, this work incorporates the sensitivity of the neural network as described in \cite{NN_sensitivity_cite}. The updated LQR state-space matrices are calculated as follows.

\begin{equation*}
\resizebox{0.5\textwidth}{!}{%
  $\mathbf{A} = 
  \begin{bmatrix}
    0 & 1 & 0 & 0 &  0 & 0\\
    \frac{\partial \pi_1(\mathbf{x,u})}{\partial x} & \frac{\partial \pi_1(\mathbf{x,u})}{\partial v_x} & \frac{\partial \pi_1(\mathbf{x,u})}{\partial y} & \frac{\partial \pi_1(\mathbf{x,u})}{\partial v_y} & -g+\frac{\partial \pi_1(\mathbf{x,u})}{\partial \theta} & \frac{\partial \pi_1(\mathbf{x,u})}{\partial \omega}\\
    0 & 0 & 0 & 1 &  0 & 0\\
    \frac{\partial \pi_2(\mathbf{x,u})}{\partial x} & \frac{\partial \pi_2(\mathbf{x,u})}{\partial v_x} & \frac{\partial \pi_2(\mathbf{x,u})}{\partial y} & \frac{\partial \pi_2(\mathbf{x,u})}{\partial v_y} &  \frac{\partial \pi_2(\mathbf{x,u})}{\partial \theta} & \frac{\partial \pi_2(\mathbf{x,u})}{\partial \omega}\\
    0 & 0 & 0 & 0 &  0 & 1\\
    0 & 0 & 0 & 0 &  0 & 0 
  \end{bmatrix}$%
}
\end{equation*}

\begin{equation}
\mathbf{B} = 
  \begin{bmatrix}
    0   &   0\\
    \frac{\partial \pi_1(\mathbf{x,u})}{\partial u_1}   &   \frac{\partial \pi_1(\mathbf{x,u})}{u_2}\\
    0   &   0\\
    1/m+\frac{\partial \pi_2(\mathbf{x,u})}{u_1} &  1/m+\frac{\partial \pi_2(\mathbf{x,u})}{u_2}\\
    0   &   0\\
    r/I & -r/I
  \end{bmatrix}
  \label{LQR_AB_NN}
\end{equation}

\subsection{ROA}

The ROA for an equilibrium point fundamentally defines the collection of all initial states from which a dynamic system will eventually converge to this equilibrium as time approaches infinity \cite{ROA_cite}. For an equilibrium point that is locally attractive, identified as $\mathbf{x}^*$, the ROA corresponding to $\mathbf{x}^*$ is described as the most extensive subset $\mathcal{D}$ within the state space $\mathcal{X}$. This set is characterized such that if an initial condition $\mathbf{x}(0)$ belongs to $\mathcal{D}$, it guarantees that $\lim_{t\to\infty} \mathbf{x}(t) = \mathbf{x}^*$ \cite{underactuated_cite}.

This concept is critical in understanding how different starting conditions affect the long-term behavior of a system. In practical terms, knowing the ROA helps in determining the robustness of a system's stability—larger ROAs suggest that the system can maintain stability across a broader range of initial conditions, enhancing its reliability and safety.

\subsection{Lyapunov Analysis}

A Lyapunov function, $V(\mathbf{x})$, is a scalar function that can be employed to assess the stability of an equilibrium point, $\mathbf{x}=\mathbf{0}$, of a dynamical system \cite{lyap_cite}. It is required that this function is positive definite and its derivative is negative definite. These functions provide a method to prove the stability of an equilibrium point of a system without solving the system's differential equations directly.

\begin{definition}
For a system described by $\mathbf{\dot{x}}=f(\mathbf{x})$, where $f$ is continuous, and within a particular region $\mathcal{D}$ surrounding the origin (an open subset of $\mathbb{R}^n$ that includes the origin), if a scalar, continuously differentiable function $V(\mathbf{x})$ can be generated, such that


\begin{equation}
\begin{aligned}
    V(\mathbf{x}) &\succ 0, \\
    \dot{V}(\mathbf{x}) &= \frac{\partial V}{\partial \mathbf{x}} f(\mathbf{x}) \prec 0, \mbox{ and} \\
    V(\mathbf{x}) &\to \infty \mbox{ whenever } \|\mathbf{x}\| \to \infty
\end{aligned}
\label{lyap_def}
\end{equation} then the origin, $\mathbf{x}=\mathbf{0}$, is globally asymptotically stable. 

The level sets of a Lyapunov function, which are the sets where $V(\mathbf{x})=c$ for a constant $c$, can be used to estimate the boundaries of the ROA.

\end{definition}

\begin{figure*}[!t]
  \begin{center}
    \begin{subfigure}[b]{0.45\textwidth}
      \centering
      \includegraphics[height=2.7in]{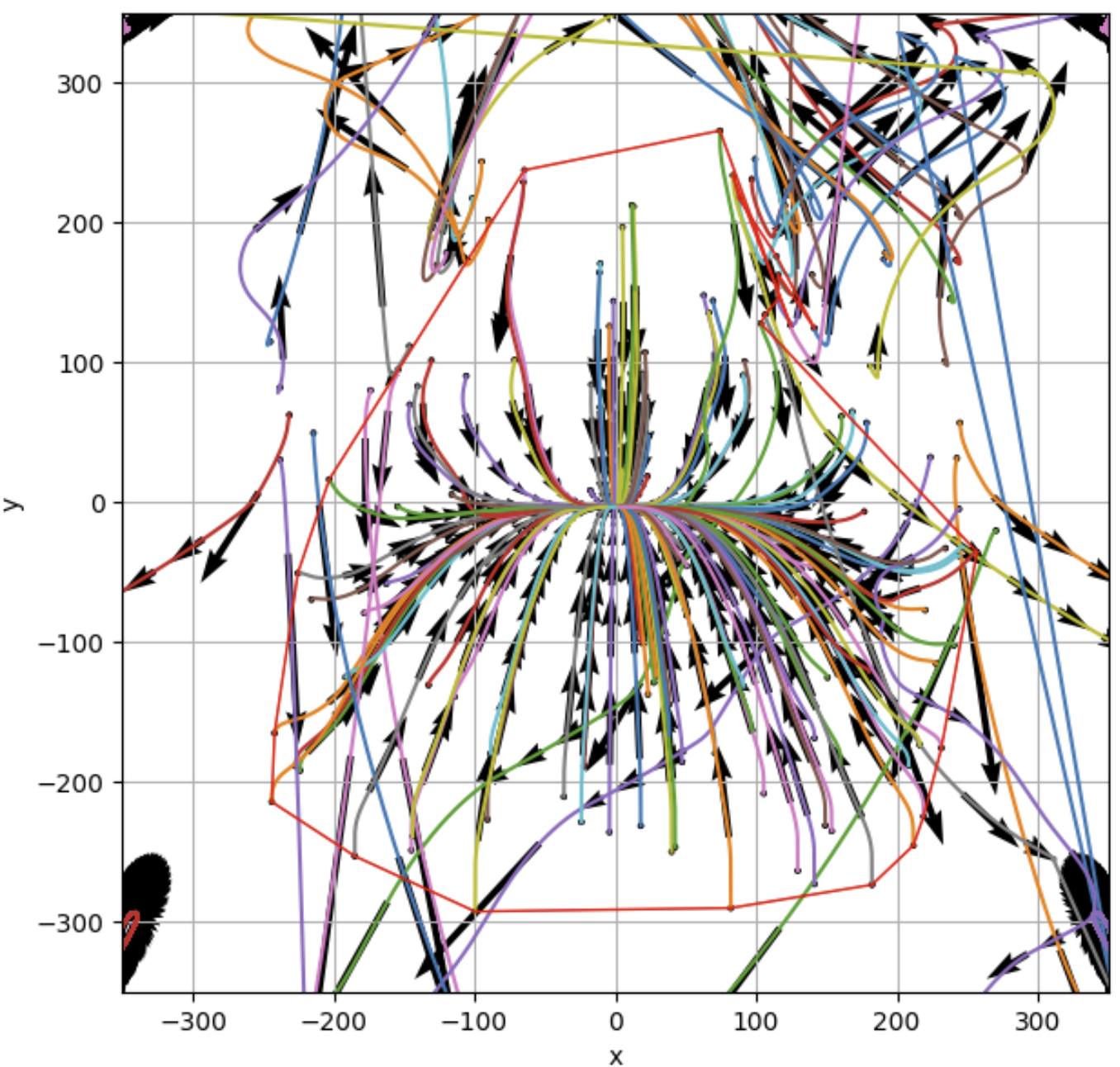}
      \caption{ROA of planar quadrotor without disturbance estimation}
      \label{ROA_no_NN}
    \end{subfigure}
    \qquad
    \begin{subfigure}[b]{0.45\textwidth}
      \centering
      \includegraphics[height=2.7in]{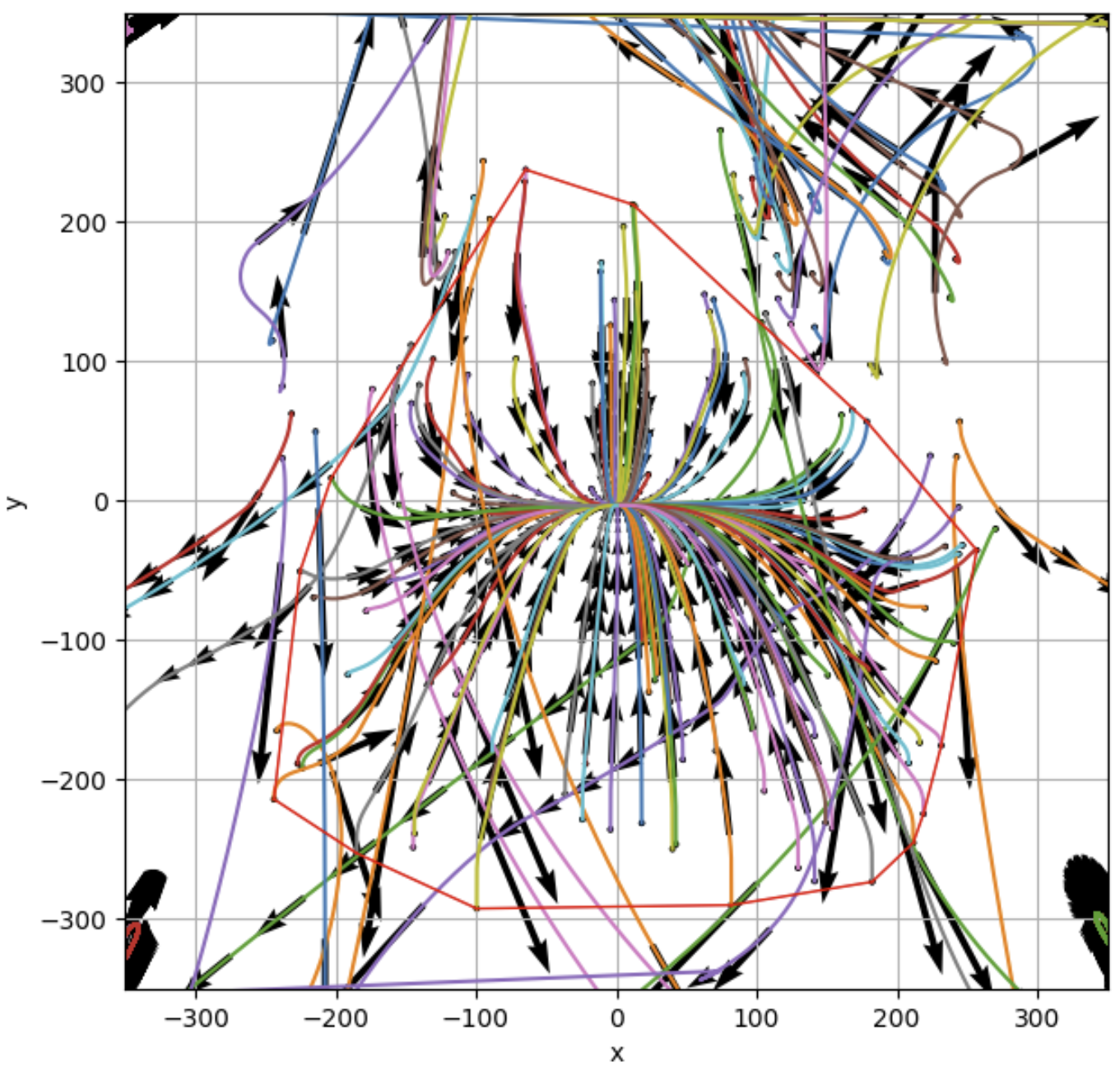}
      \caption{ROA of planar quadrotor with disturbance estimation}
      \label{ROA_NN}
    \end{subfigure}
  \end{center}
  \caption{Comparison of the ROA of the planar quadrotor with and without disturbance estimation - ROA of the convex hull is outlined in red. Each color denotes the trajectory of each initial point towards $(0,0)$; arrows define velocities at each trajectory point.}
  \label{fig:ROA_comparison}
\end{figure*}

\section{SIMULATIONS AND VALIDATION}



\subsection{Simulation Setup}

The planar quadrotor equipped with the LQR controller was subjected to simulations where $150$ random initial points, ranging from $-250\,\mathrm{m}$ to $250\,\mathrm{m}$, were steered towards the origin, $(0,0)$. These points were carefully chosen to ensure an adequate sample size for both ROA comparison and neural network training purposes. 

The LQR controller, which is the control principle used in this work was defined as in \eqref{LQR_state_space}. It generates the desired thrust, $\mathbf{u}_k$ for each rotor at the time step, $k$, using the desired state, $\mathbf{x}_{des}$, and the current state at time step $k$, $\mathbf{x}_k$. The weighting matrices, $\mathbf{Q}$ and $\mathbf{R}$, of the LQR controller were tuned, and the optimal control gain, $\mathbf{K}$, was calculated as outlined in \eqref{LQR_K}. The controller gains are adjusted, enhancing the precision and operational efficiency of the quadrotor as it navigates toward its designated target.


\begin{equation}
  \mathbf{Q} = \mathrm{diag}(1, 100, 10,  100, 10, 1 000)
\label{LQR_Q}
\end{equation}

\begin{equation}
  \mathbf{R} = \mathrm{diag}(10 000, 10 000)
\label{LQR_R}
\end{equation}

\begin{equation}
  \mathbf{K} = 
  \begin{bmatrix}
    -0.007 & -0.076 & 0.022 & 0.125 & 0.604 & 0.243 \\
    0.007 & 0.076 & 0.022 & 0.125 & -0.604 & -0.243
  \end{bmatrix}
\label{LQR_K}
\end{equation}


The nominal dynamic model of the planar quadrotor was formulated based on the principles of physics. It defines the planar quadrotor’s state-space model in \eqref{dynamics}. The trigonometric terms, $\sin(\theta)$, and $\cos(\theta)$ were approximated using the 3rd-order Taylor series expansion. A simulated disturbance model in \eqref{disturbance} was integrated into the nominal dynamic model to replicate real-world conditions in the simulated planar quadrotor plant. At each $k^{th}$ time step, it calculates the disturbance, $\pi(\mathbf{x},\mathbf{u})_k$ and integrates into the nominal dynamic model. The state derivatives, $\mathbf{\dot{x}}_k$ are computed using the states, $\mathbf{x}_k$ and the control input, $\mathbf{u}_k$ at the $k^{th}$ time step.

During the simulation, some of the initial points successfully converged to the origin, whereas others diverged considerably from it. The ROA was identified by constructing a convex hull around the furthest set of stable points, outlining the outer boundary of the area where the stable initial points converge to the equilibrium safely. The ROA was computed both with and without including the estimated disturbance model, allowing for a comparative analysis of the results. 


\subsection{ROA with Nominal Dynamic Model}

From the previously generated set of $150$ initial points, $100$ were selected for this process. These points were subjected to simulations directing them toward the equilibrium point at $(0,0)$. Once the result was generated, additional $50$ points closer to the boundary were added to this set to further expand the ROA. Then, the ROA was generated as the convex hull, and the area of the convex hull was calculated. The resultant ROA of this approach is depicted in Figure \ref{ROA_no_NN}, with the total area calculated to be $81,763.019\,\mathrm{m^2}$. 
 



\begin{figure}[!b]
  \centering
  \includegraphics[height=1.6in]{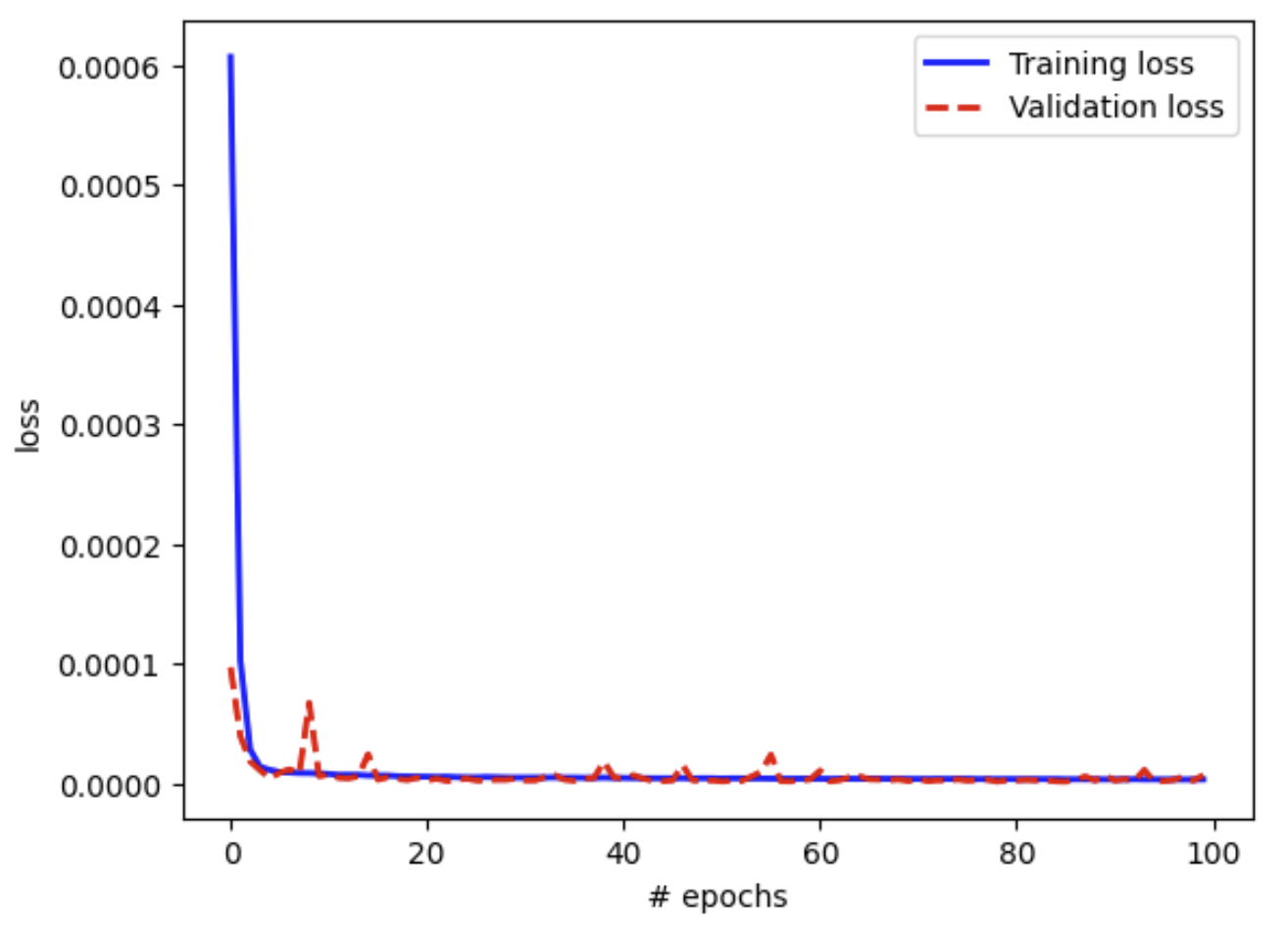}\\
  \caption{Training and validation loss}
\label{NN_loss}
\end{figure}

\subsection{ROA with Learnt Dynamic Model}

Building upon the foundational ROA assessment, this section explores the enhancement of the ROA by incorporating estimated disturbances via a neural network. This neural network consists of two fully connected hidden layers with $30$ and $15$ neurons, respectively, with $\mathrm{tanh}$ as the activation function. Inputs to the network were chosen as, $^bv_x$, $^bv_y$, $\theta$, $u_1$, and $u_2$.

A dataset consisting of $50$ stable trajectories generated from the simulations was employed to train the neural network. $20\%$ of this dataset was randomly chosen as the validation in each epoch, with the remainder being used for the training process. The neural network underwent training over $100$ epochs, during which the training and validation losses were computed at each epoch. The trained model demonstrated high accuracy with an average validation loss of $10^{-6}$. The training and validation losses for each epoch are depicted in Figure \ref{NN_loss}, illustrating the effectiveness of the neural network in capturing and predicting dynamic disturbances.

To validate the effectiveness of the learned disturbances, the same set of $169$ initial points previously used in simulations without the neural network's predictions were again simulated, directing them towards the same equilibrium point at $(0,0)$. Disturbances predicted by the trained neural network were then integrated into the quadrotor's dynamics and the LQR state-space matrices $\mathbf{A}$ and $\mathbf{B}$ were updated as in \eqref{LQR_AB_NN} to assess their impact.

Following the integration of these predicted disturbances, the ROA was regenerated using the convex hull method, as was done in previous simulations. This method involved plotting the outermost stable points that the quadrotor could reach, forming a convex boundary. The results of this updated simulation are illustrated in Fig. \ref{ROA_NN}, where the area of the ROA was calculated to be $88,426.015\,\mathrm{m^2}$. Notably, this area is significantly larger than the area calculated without these disturbances. Comparing Fig. \ref{ROA_no_NN} and Fig. \ref{ROA_NN}, a few points that were unstable before have become stabilized at the equilibrium point, which caused a larger ROA.

\begin{figure}[!t]
  \centering
  \includegraphics[height=1.5in]{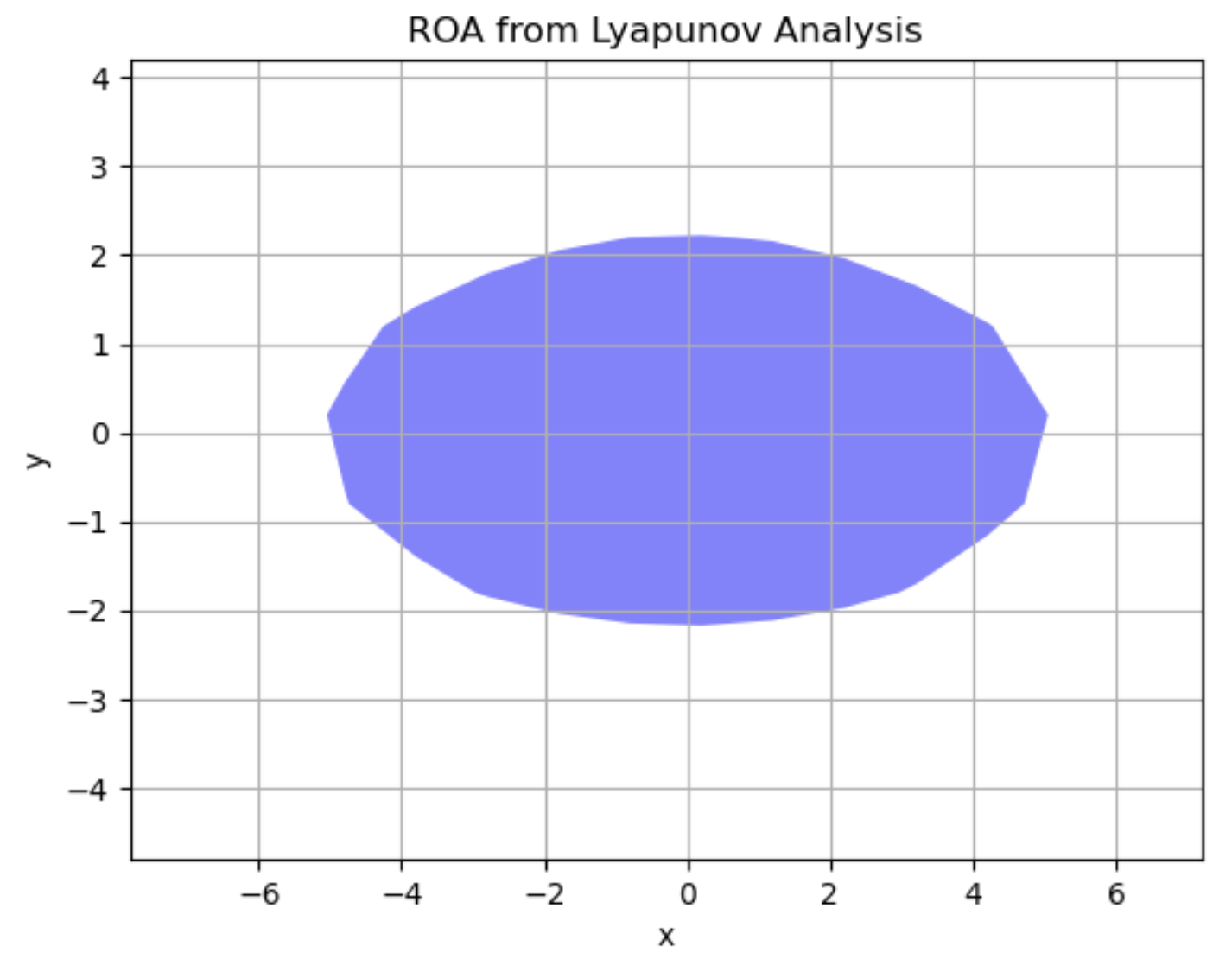}\\
  \caption{ROA of planar quadrotor from Lyapunov Analysis}
\label{ROA_Lyap}
\end{figure}

\subsection{Lyapunov Analysis}

The results of this proposed method were then compared against those obtained from Lyapunov analysis to illustrate the conservatism of the latter. In this approach, the ROA is calculated from SOS optimization, using a level set of a Lyapunov function. The Lyapunov function was defined as the quadratic cost-to-go function associated with the LQR controller, as shown in \eqref{lyap} where $\mathbf{S}$ is the solution associated with algebraic Riccati equation \cite{underactuated_cite}. The SOS optimization was applied by incorporating an SOS constraint into the model to confirm this Lyapunov function is within the ROA. The ROA generated by this method is shown in Fig. \ref{ROA_Lyap}. It is evident that the area from the Lyapunov analysis, measuring $12.715\,\mathrm{m^2}$, is substantially smaller than the true ROA.

\begin{equation}
  V(\mathbf{x}) = \mathbf{x}^T\mathbf{S}\mathbf{x},
\label{lyap}
\end{equation} where, $\mathbf{S}$ is the solution associated algebraic Riccati equation \cite{underactuated_cite}.

A comprehensive comparison of areas of the ROA, including those estimated by Lyapunov analysis, without neural network intervention, and with neural network integration, is presented in Table \ref{comparison}. This table facilitates a direct visual comparison of how each model affects the size and stability region of the ROA.

\begin{table}[!t]
    \caption{ROA area comparison}
    \label{table_example}
    \begin{center}
    \begin{tabular}{|c|c|}
    \hline
    \thead{ROA computation technique} & \thead{Area} $(\,\mathrm{\mathbf{m^2}})$ \\
    \hline 
    Graphical method with disturbance estimation & 88,426.0145 \\
    \hline
    Graphical method without disturbance estimation & 81,763.019 \\
    \hline
    Lyapunov analysis & 12.715 \\
    \hline
    \end{tabular}
    \end{center}
\label{comparison}
\end{table}

\section{CONCLUSIONS}

This work investigates enhancing the ROA in systems governed by an LQR control policy through improvements in the dynamic model facilitated by machine learning. Specifically, neural networks are utilized to identify and learn the unknown disturbances, a technique that shifts away from traditional dynamic modeling to address the complexities and unpredictabilities of real-world scenarios. By incorporating these disturbances into the dynamic model, the ROA is notably enhanced, illustrating the benefits of merging machine learning with control theory to develop more robust and adaptable control systems. These improvements significantly improve the system's stability and safety, highlighting the method's wider relevance for augmenting the performance of autonomous systems facing unpredictable conditions. Additionally, the expanded ROA demonstrates the neural network's capability to improve the system's resilience to dynamic variations, thus enhancing operational safety and robustness. These results emphasize the importance of applying sophisticated machine-learning strategies in control systems to elevate their functionality and safety across diverse operational environments.

The scope of this research is expanding to integrate the methodology into 3D quadrotor systems, which are currently underway. By applying the learned insights and methodologies to 3D quadrotors, the research aims to tackle the additional challenges presented by these more complex systems.



\addtolength{\textheight}{-12cm}   









\bibliographystyle{IEEEtran}
\bibliography{IEEEabrv,references}

\end{document}